\documentclass[twocolumn, trackchanges]{aastex62}

\received{2019 March 30}
\revised{2019 May 10}
\accepted{2019 May 12}
\submitjournal{ApJL}

\shortauthors{Motogi et al.}

\begin{document}

\title{The first bird's-eye view of a gravitationally unstable accretion disk in high-mass star formation}

\correspondingauthor{Kazuhito Motogi}
\email{kmotogi@yamaguchi-u.ac.jp}

\author{Kazuhito Motogi }
\affil{Graduate School of Sciences and Technology for Innovation,Yamaguchi University, Yoshida 1677-1, Yamaguchi 753-8512, Japan}

\author{Tomoya Hirota}
\affil{Mizusawa VLBI Observatory, National Astronomical Observatory of Japan, Osawa 2-21-1, Mitaka, Tokyo 181-8588, Japan}
\affil{Department of Astronomical Sciences, SOKENDAI (The Graduate University for Advanced Studies), Osawa 2-21-1, Mitaka, Tokyo 181-8588, Japan}

\author{Masahiro N. Machida}
\affil{Department of Earth and Planetary Sciences, Faculty of Sciences, Kyushu University, Motooka 744, Nishi-ku, Fukuoka-shi, Fukuoka 819-0395, Japan}

\author{Yoshinori Yonekura}
\affil{Center for Astronomy, Ibaraki University, 2-1-1 Bunkyo, Mito, Ibaraki 310-8512, Japan}

\author{Mareki Honma}
\affil{Mizusawa VLBI Observatory, National Astronomical Observatory of Japan, Hoshigaoka 2-12, Mizusawa, Oshu, Iwate 023-0861, Japan}
\affil{Department of Astronomical Sciences, SOKENDAI (The Graduate University for Advanced Studies), Osawa 2-21-1, Mitaka, Tokyo 181-8588, Japan}

\author{Shigehisa Takakuwa}
\affil{Department of Physics and Astronomy, Graduate School of Science and Engineering, Kagoshima University, 1-21-35 Korimoto, Kagoshima, Kagoshima 890-0065, Japan}

\author{Satoki Matsushita}
\affil{Academia Sinica, Institute of Astronomy and Astrophysics (ASIAA), P.O. Box 23-141, Taipei, 10617 Taiwan}

\begin{abstract}
We report on the first bird's-eye view of the innermost accretion disk around the high-mass protostellar object G353.273+0.641, taken by Atacama Large Millimter/submillimeter Array long-baselines. 
The disk traced by dust continuum emission has a radius of 250 au,  surrounded by the infalling rotating envelope traced by thermal CH$_3$OH lines.  
This disk radius is consistent with the centrifugal radius estimated from the specific angular momentum in the envelope. 
The lower-limit envelope mass is $\sim$5-7  M$_{\sun}$ and accretion rate onto the stellar surface is 3 $\times$ 10$^{-3}$ M$_{\sun}$ yr$^{-1}$ or higher. 
The expected stellar age is well younger than 10$^{4}$ yr, indicating that the host object is one of the youngest high-mass objects at present. 
The disk mass is 2-7 M$_{\sun}$, depending on the dust opacity index. 
The estimated Toomre's $Q$ parameter is typically 1-2 and can reach 0.4 at the minimum. 
These $Q$ values clearly satisfy the classical criteria for the gravitational instability, and are consistent with the recent numerical studies. 
Observed asymmetric and clumpy structures could trace a spiral arm and/or disk fragmentation. 
We found that 70$\%$ of the angular momentum in the accretion flow could be removed via the gravitational torque in the disk. 
Our study has indicated that the dynamical nature of a self-gravitating disk could dominate the early phase of high-mass star formation. 
This is remarkably consistent with the early evolutionary scenario of a low-mass protostar. 
\end{abstract}


\section{Introduction} \label{sec:intro}
 In the last two decades, 10 or more accretion disk candidates, which play critical roles in low-mass star formation, have been found for high-mass stars \citep{2006Natur.444..703C, 2016A&ARv..24....6B}. 
The final stellar mass and evolutionary pathway of high-mass protostars should be determined by physical processes inside a disk \citep[e.g.,][]{2010ApJ...721..478H}. 
However, almost all of the known resolved disks around high-mass stars are nearly edge-on to the disk; this is easy to detect due to the maximized rotating motion along the line  of sight (LOS). 
Consequently, no one has ever directly seen what's happening at the innermost region of a disk. 
This Letter reports on the first Atacama Large Millimeter/submillimeter Array (ALMA) long baseline imaging toward the nearly face-on high-mass protostellar object G353.273+0.641 (hereafter G353).

G353 is a relatively nearby high-mass object ($\sim$1.7 kpc) without any compact HII region \citep{2016PASJ...68...69M, 2017ApJ...849...23M}. 
The stellar mass is $\sim$10 M$_{\sun}$ and is still in the main accretion phase, as suggested by active outflow activities \citep{2013MNRAS.428..349M, 2016PASJ...68...69M}. 
Previous studies have shown that G353 has an almost face-on accretion system, where the disk rotational axis is inclined $\sim$ 8\degr from the LOS \citep{2016PASJ...68...69M, 2017ApJ...849...23M}. 
Therefore, G353 is the best target for direct imaging of any radial structures in the disk, minimizing a self-absorption effect by an optically thick dusty system as reported by recent ALMA observations \citep[e.g.,][]{2017A&A...603A..10B}. 
In particular, \citet{2017ApJ...849...23M} observed 6.7 GHz CH$_{3}$OH masers and found the three-dimensional infalling streams from 100 to 15 au. 
They discussed possibilities of very small initial angular momentum and/or significant angular momentum transfer outside 100 au in radius. 
This work directly resolved 10$^{2}$-10$^{3}$ au scale and examined these possibilities.

\section{Observation and reduction} \label{sec:obs}
We performed high-resolution ($\sim$85 au) imaging of the dust continuum and several CH$_3$OH lines at 150 GHz in ALMA Cycle 4.   
Forty-five antennas were used in the long baseline configuration (up to $\sim$12 km). 
The total on-source time was 5 minutes. 
We observed four 1.875 GHz wide sub-bands centered on  142.692, 144.571, 154.613, and 156.491 GHz (ALMA band 4). 
Only the sub-band at 156.491 GHz, which had the highest spectral resolution of 488.281 kHz (0.94 km s$^{-1}$), was used for the line analysis. 
We integrated all the four sub-bands for the dust continuum analysis after the line subtraction. 

The data were calibrated in the standard manner of ALMA by using the CASA software package version 5.3.0 \citep{2007ASPC..376..127M}. 
No self-calibration was performed.  
The uncertainties are dominated by the error of the absolute flux scaling using the quasar J1733--1304 ($\sim$10\%). 
Therefore we adopt 10\%  error for all the fluxes and/or brightness temperature ($T_{\rm B}$) in this Letter. 

Original synthesized beams were 0\arcsec.085$\times$0\arcsec.050 (145$\times$85 au$^2$) and 0\arcsec.099$\times$0\arcsec.075 (168$\times$128 au$^2$) for the continuum and CH$_{3}$OH lines, respectively. 
The beam position angle for the continuum and lines were 48\degr.7 and 51\degr.4 (east of north), respectively. 
For fair comparison between new 150 GHz data and previous J-VLA 45 GHz data, the obtained continuum image was finally reconstructed by using the circular beam of 0\arcsec.050 $\times$ 0\arcsec.050 as in the previous study \citep{2017ApJ...849...23M}, 
where all the clean components were reconvolved by the circular beam in the deconvolution process. 
The final image noise level (1$\sigma$) is 2.7 K for the continuum, and 7.5 K for line emission after the line stacking analysis (see below). 
The continuum peak position was 17$^{\rm h}$26$^{\rm m}$01$^{\rm s}$.58798, -34\degr15\arcmin14\arcsec.9175 (J2000.0). 
We used this position as the coordinate origin of all the spatial maps in this Letter. 

\section{Results and discussions} \label{sec:result}
\subsection{Dust continuum} \label{sec:dust}
Fig. 1a shows the dust continuum image of the accretion disk that consists of the compact central emission and resolved structure. 
The latter was highlighted in Figure 1b, where the former compact emission was subtracted by elliptical Gaussian fitting (Appendix \ref{app:dust}). 
The compact emission has an averaged diameter of $\sim$160 au (Table \ref{tab1}). 
The peak $T_{\rm B}$ is 480 K. 
The resolved structure is almost “ring-like” with the outer radius of 250 au. 
There are an arc-like elongated feature in the east and several clumpy features in the west. 
The peak $T_{\rm B}$ in the east side (up to 126 K) is 2-3 times brighter compared to the other, less bright parts. 
This asymmetry would reflect the contrast of the surface density rather than the asymmetric temperature profile, as the disk temperature should be basically dominated by stellar radiation in this small scale. 
It should be noted that, if the disk actually has a non-axisymmetric structure such as spiral arms, the dynamical heating effect can cause a deviation from the axisymmetric temperature profile. 
Although few limited studies observed the innermost accretion system in high-mass star-formation \citep{2014A&A...566A..73C, 2017A&A...603A..10B, 2017NatAs...1E.146H}, 
this is the first completely resolved view of such a compact non-axisymmetric structure inside a disk. 

The compact emission is consistent with the continuum emission at  45 GHz previously reported in \citet{2017ApJ...849...23M} as shown in Fig. 1b and Table \ref{tab1}, 
indicating that we could successfully subtract the hot and compact emission in the innermost region. 
The spectral index is 2.5 ± 0.2 between 45 and 150 GHz, considering the flux uncertainty of 10\%. 
The optical depth, averaged surface density ($N_{\rm H2}$), and total mass of the compact emission are 2.0$^{+1.4}_{-0.9}$, 8.2$^{+5.7}_{-3.8}$$\times$10$^{25}$ cm$^{-2}$ and 0.8$^{+0.5}_{-0.4}$ M$_{\odot}$, respectively (Appendix \ref{app:dust}). 
On the other hand, the optical depth in the resolved structure was determined as 0.68$^{+0.36}_{-0.30}$ at the eastern peak and 0.31$\pm$0.12 on average, 
which corresponds to the surface densities of 2.8$^{+1.5}_{-1.2}$$\times$10$^{25}$ cm$^{-2}$ and 1.3$\pm$0.5$\times$10$^{25}$ cm$^{-2}$, respectively. 
The total mass of the dusty accretion system is 1.9$^{+1.0}_{-0.8}$ M$_{\odot}$, including the compact component. 
Here we adopted the dust mass opacity of 0.59  cm$^2$g$^{-1}$ at 150 GHz and the gas-to-dust ratio of 100 (Appendix \ref{app:dust}). 

Fig. 1c shows the entire continuum emission in a further large-scale envelope ($\sim$700 au in radius). 
The integrated flux density is 316 mJy ($>$ 3$\sigma$ level), including 140 mJy from the compact emission within 250 au.  
The mass outside of 250 au is $\sim$5 M$_{\sun}$, which is estimated by the graybody emission adopting the same dust parameters and the averaged temperature of $\sim$100 K that is deduced from line emissions (see the next subsection). 
The total mass of the entire system within 700 au radius is 7 M$_{\sun}$.  
The uncertainty of these mass estimations originates primarily from the uncertainty of the dust parameter. 
This will be discussed later. 

\begin{figure*}[!htb]
\epsscale{1.2}
\plotone{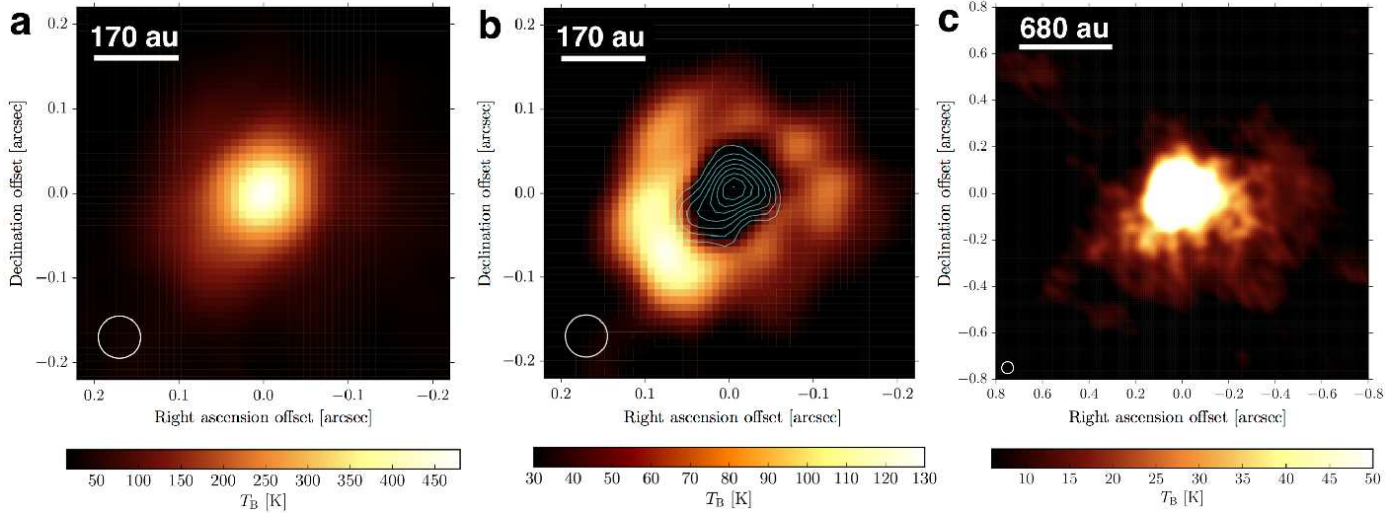}
\caption{(a): Original dust continuum image near the center.  
The white circle represents the synthesized beam for continuum. 
(b): Residual image after the subtraction of the compact emission. Contours show the continuum at 45 GHz in \citet{2017ApJ...849...23M}, which are every 25 K starting from 100 K (5$\sigma$). 
(c): The entire distribution of dust continuum emission. The color scale is artificially saturated at 50 K (18.5$\sigma$), in order to highlight diffuse extended emission. 
\label{fig:1}}
\end{figure*}

\begin{deluxetable*}{cccccc}
\tablecaption{The elliptical Gaussian parameters of the compact component. \label{tab1}}
\tablewidth{0pt}
\tablehead{
Frequency & Major Axis & Minor Axis & Position Angle & Total Flux & Peak $T_{\rm B}$ \\
(GHz) & \multicolumn2c{(au)} & (degree) &  (mJy) & (K) 
}
\decimals
\startdata
45$^{a}$ & 209 $\pm$ 7 & 124 $\pm$ 14 & 143 $\pm$ 5 & 3.5 $\pm$ 0.1 & 278  \\ \hline
150 & 177 $\pm$ 0.3 & 136 $\pm$ 0.2 & 149 $\pm$ 0.3 & 72.5 $\pm$ 0.2 & 480  \\ \hline
\enddata
\tablenotetext{a}{The data at 45 GHz is from \citet{2017ApJ...849...23M}}
\end{deluxetable*}

\subsection{CH$_3$OH lines} \label{sec:ch3oh}
We have detected seven CH$_{3}$OH lines (Fig. 2a and Table \ref{tab2}). 
Fig. 2b presents the integrated intensity map of the molecular envelope that has a comparable scale with that of the entire dust emission. 
Here, we stacked three CH$_3$OH lines, with the upper-state rotational energy higher than 70 K (Appendix \ref{app:ch3oh}). 
It should be noted that all of the detected lines are thought to be not maser but thermal emission, because of the extended structure and moderate brightness. 
The envelope size is 700 au in radius and completely encloses the compact dusty structure. 
The CH$_3$OH emission has been significantly decreased toward the center, which is consistent with the optically-thick dust emission. 
Almost all of the CH$_{3}$OH emission is from beyond 250 au in radius. 
Fig. 2c shows the position-velocity ($PV$) diagram of the CH$_3$OH emission along the north--south direction (Appendix \ref{app:ch3oh}). 
The entire velocity structure is roughly symmetric against the systemic velocity of the natal cloud ($\sim$-4.5 km s$^{-1}$). 

We detected both the rotational spin-up motion and infalling motion at the same time \citep[e.g.,][]{2014Natur.507...78S, 2017A&A...603A..10B}. 
These kinematic features could be basically reproduced by a simple model of the infalling rotating envelope, as in the case of low-mass protostars \citep{2014Natur.507...78S, 2016ApJ...824...88O}. 
Fig. 2d shows the schematic view of our model, which is described in Appendix \ref{app:ch3oh}. 
The envelope is rotating counterclockwise around the rotational axis that is inclined 8$^{\circ}$ from the LOS \citep{2016PASJ...68...69M, 2017ApJ...849...23M}. 
The envelope becomes thicker outward with a constant flaring angle. We adopted the specific angular momentum of 2.4$\times$10$^{21}$ cm$^{2}$ s$^{-1}$ in the model. 
This corresponds to the centrifugal radius of $\sim$250 au consistent with the outer radius of the resolved structure in Fig. 1b, assuming the stellar mass of 10 M$_{\sun}$. 

We calculated the line-of-sight velocities on the midplane and the near/far side of the envelope surface (Fig. 2d). 
We have found that the envelope should have a relatively wide flaring angle ($\sim$ 35$^{\circ}$), 
in order to reproduce the rotational spin-up motion and the centrifugal brake against the infall acceleration. 
For example, if the envelope is geometrically thin with a much smaller flaring angle, the velocity field coincides with that in the midplane. 
On the other hand, the specific angular momentum that is significantly smaller than the assumed value could not reproduce the feature of the centrifugal braking. 
Conversely, the rotational spin-up feature at the angular offset of $\pm$ 0$\arcsec$.2 could not be reproduced by the significantly larger angular momentum. 
Therefore, we suggest that the inner dust emission traces a centrifugal disk and the rotating accretion flow reaches Keplerian rotation inside 250 au. 
It should be noted that further fine-tuned model fitting is beyond the scope of this study, 
because we clearly require higher spectral resolution and better sensitivity to detect any faint high-velocity emission. 

We also detected four $^{13}$CH$_3$OH lines (Fig. 2a and Table \ref{tab2}). 
The spatial structures of those $^{13}$CH$_3$OH lines are almost identical to that of the main isotopes. 
Surprisingly, the observed line ratio (CH$_3$OH/$^{13}$CH$_3$OH) is very small ($\sim$ 1.3), suggesting that these line emissions are optically thick and almost thermalized. 
The averaged $T_{\rm B}$ of these optically thick CH$_{3}$OH lines is $\sim$100 K. 
We used this value as the dust temperature in the outer envelope. 

If we adopt the isotopic abundance ratio $^{12}$C/$^{13}$C $\sim$ 60, the averaged optical depth of $\sim$ 90 is expected for the main isotopes. 
By using the non-local thermodynamic equilibrium radiative transfer code \citep[RADEX:][]{2007A&A...468..627V}, we estimated the lower-limit density ($n_{\rm H2}$) and column density ($N_{\rm CH3OH}$) as 3$\times$10$^9$ cm$^{-3}$ and 10$^{19}$ cm$^{-2}$, respectively. 
Here, we adopted a kinematic temperature of 100 K and velocity dispersion of 5 km s$^{-1}$, which is the observed line width (FWHM). 
The lower-limit $N_{\rm H2}$ is $\sim$ 10$^{25}$ cm$^{-2}$, assuming the relative CH$_3$OH abundance of 10$^{-6}$,  which is the highest abundance observed in hot cores \citep[e.g.,][]{2007A&A...465..913B}. 
We finally obtained a total envelope mass of 7 M$_{\sun}$ using the averaged radius of 700 au. 
This is comparable with that estimated from the dust emission ($\sim$ 5 M$_{\sun}$). 

We note that actual dust temperature might be colder than the gas temperature estimated by line emission, because of different spatial locations in the disk/envelope as suggested in recent observational studies \citep[e.g.,][]{2017A&A...603A..10B}. 
The envelope mass estimated by dust emission could be larger in this case.

\begin{figure*}[htb]
\epsscale{1.2}
\plotone{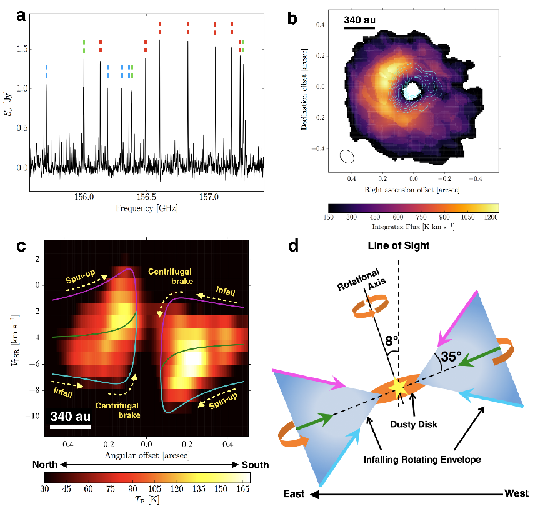}
\caption{(a): Spectrum of the detected CH$_{3}$OH lines. We marked several strong emission by dashed lines. Red and blue dashed lines indicate CH$_{3}$OH and $^{13}$CH$_{3}$OH transitions, respectively. 
Green dashed lines indicate blended lines, where multiple CH$_{3}$OH or $^{13}$CH$_{3}$OH transitions are blended. 
(b): Integrated flux map of the CH$_3$OH emission (Appendix \ref{app:ch3oh}). 
Contours show the dust emission in Fig. 1b at every 15 K starting from 40 K (15$\sigma$). 
The black ellipse represents the synthesized beam for line emissions. 
(c) $PV$ diagram of the stacked CH$_{3}$OH emission along $x$ = 0$\arcsec$ in Fig. 2b. 
$V_{\rm LSR}$ denotes the LOS velocity with respect to the local standard of rest. 
The origin of the angular offset is the continuum peak position. 
Magenta, cyan, and green lines indicate the LOS velocities on the near/far side and midplane calculated by the envelope model as shown in  Fig. 2d, respectively (Appendix \ref{app:ch3oh}). 
(d): Schematic view of the system seen from edge-on. 
Magenta, cyan, and green arrows indicate infalling directions, corresponding to the velocity models in Fig. 2c. 
\label{fig:2}}
\end{figure*}

\begin{deluxetable*}{ccccc}
\tablecaption{Detected CH$_3$OH and $^{13}$CH$_3$OH lines. \label{tab2}}
\tablewidth{0pt}
\tablehead{
Transition$^{a}$ & Frequency & Upper-state Energy & Ortho/Para$^{\dagger}$ &Category$^{\ddagger}$ \\
 $J_{K}$& (GHz) & (K) & & 
}
\decimals
\startdata
\multicolumn5c{Stacked CH$_{3} $OH Lines} \\ \hline 
$2_1$ -- $3_0$ & 156.60235 & 21.4 &A&Cold \\ \hline
$6_2$ -- $7_1$& 156.12770 & 86.5 &A& Hot \\ \hline
$8_0$ -- $8_{-1}$& 156.48886 & 96.6&E& Hot \\ \hline
$7_0$ -- $7_{-1}$ & 156.82848 & 78.1 &E& Hot \\ \hline
$6_0$ -- $6_{-1}$ & 157.04859 & 61.8 &E& Cold \\ \hline
$5_0$ -- $5_{-1}$ & 157.17896 & 47.9 &E& Cold \\ \hline
$4_0$ -- $4_{-1}$ & 157.24604& 36.3 &E& Cold \\ \hline
\multicolumn5c{Detected $^{13}$CH$_{3} $OH Lines} \\ \hline 
 $8_0$ -- $8_{-1}$& 155.69574 & 94.6 &E& -- \\ \hline
 $6_0$ -- $6_{-1}$& 156.18652 & 60.7 &E& -- \\ \hline
 $5_0$ -- $5_{-1}$& 156.29937 & 47.1 &E& -- \\ \hline
 $4_0$ -- $4_{-1}$& 156.35640 & 35.8 &E& -- \\ \hline
\enddata
\tablenotetext{a}{The rotational energy is denoted $J_{K}$, where $J$ and $K$ indicate the rotational quantum number and its projection along the symmetry axis, respectively. }
\tablenotetext{\dagger}{Nuclear spin state of three hydrogens around the carbon (E: Ortho, A: Para). }
\tablenotetext{\ddagger}{We made a stacking image only for the hot transitions (Fig. 2b), avoiding foreground absorption via the blue-shifted outflow lobe. }
\end{deluxetable*}

\subsection{Accretion rate and stellar properties} \label{sec:mdot}
The kinematic model indicates that an infalling timescale of the molecular envelope is $\sim$ 10$^3$ yr.  
This was estimated by numerically solving the equation of motion along the envelope surface for the mass particle at the outer edge. 
Here we only considered the stellar gravity and centrifugal force in the equation. 
On the other hand, the averaged envelope density of 3.5 $\times$ 10$^{-15}$ g cm$^{-3}$, where 6 $M_{\sun}$ is divided by the volume of the modeled envelope, corresponds the freefall time of $\sim$ 1.1 $\times$ 10$^{3}$ yr. 
These are comparable, but the numerical value is slightly shorter than the freefall time, because of the stellar gravity. 
The averaged infall rate inside the envelope is $\sim$ 6 $\times$ 10$^{-3}$ M$_{\odot}$ yr$^{-1}$. 
We expect an effective accretion rate onto the stellar surface as 3$\times$ 10$^{-3}$ M$_{\odot}$ yr$^{-1}$ or higher, 
because 50\% of accreting mass could be converted to the outflow in case of high-mass star-formation \citep[e.g.,][]{2017MNRAS.470.1026M}. 

This infall rate is clearly larger than that estimated from the Jeans mass (i.e., $\sim$ $c^{3}_s/G$). 
It becomes $\sim$ 10$^{-4}$ M$_{\odot}$ yr$^{-1}$, even adopting the current envelope temperature of 100 K. 
This discrepancy is usual case in high mass star-formation, 
where we must consider the so-called effective Jeans mass, including a non-thermal velocity component such as turbulent or Alfv$\acute{\rm e}$n velocities \citep[e.g.,][]{2009ApJ...691..823H, 2013ApJ...774L..31I}. 
We can replace $c_s$ with $c_{\rm eff}$, which is the root sum square of these velocity components in this case. 
The observed infall rate corresponds $c_{\rm eff}$ of $\sim$3 km s$^{-1}$, suggesting a highly turbulent and/or strongly magnetized condition of the initial core. 
Such a condition may be achieved by dynamical compression via cloud--cloud collision \citep[e.g.,][]{2013ApJ...774L..31I}, rather than simple gravitational condensation. 

The age of G353 could be only 3$\times$10$^{3}$ yr, considering the infall rate of 3$\times$ 10$^{-3}$ M$_{\odot}$ yr$^{-1}$. 
We note that the accretion rate onto the stellar surface from the disk can be highly variable in the case of the gravitationally unstable disk; however, 
the time-averaged accretion rate should be consistent with the infall rate onto the disk from the envelope \citep[e.g.,][]{2018MNRAS.473.3615M} after subtracting the outflow rate. 
We suggest that G353 is one of the youngest ($<$ 10$^4$ yr) high-mass objects, currently known, even considering an overall error of factor 3. 
The final stellar mass could be 13-17 M$_{\sun}$ depending on the outflow efficiency. 
It should be noted that this is still the lower-limit mass, as any envelope structure larger than 0$\arcsec$.5 should be resolved out in our data. 
Further observations on a larger scale are required to confirm the final stellar mass and its evolutionary stage.

\subsection{Gravitational stability of the disk} \label{sec:Q}
The disk mass within 250 au reaches almost 20\% of the stellar mass; therefore, the disk could be self-gravitating \citep{2016MNRAS.463..957F}. 
Fig. 3a shows the spatial distribution of the Toomre's $Q$ parameter \citep[][see Appendix \ref{app:Q}]{1964ApJ...139.1217T}, which is a well-known measure of the gravitational stability ($Q$ $<$ 1: the disk immediately collapses; 1 $<$ $Q$ $<$ 2: the disk is unstable against non-axisymmetric perturbations). 
The central region shows $Q$ $>$ 2 suggesting that the disk is stabilized by the stellar radiation and stronger gravity of the central star. 
On the other hand, a significant fraction of the outer disk shows that $Q$ $<$ 2 and the minimum $Q$ reaches 0.9 around the eastern local peak. 
This fact indicates that the disk is, at least, unstable against non-axisymmetric perturbations such as a spiral arm \citep[e.g.,][]{2017ApJ...835L..11T}. 
Furthermore, the averaged $Q$ of $\sim$ 1.9 outside 80 au in radius suggests that the entire disk is moderately unstable, avoiding the immediate collapse during the freefall time ($\sim$a few hundred yr). 
Such a short time scale is too short to be observed. 

Although these criteria are valid for low-mass star-formation, it may not be the case for high-mass star formation. 
Recent high-resolution numerical studies in \citet{2018MNRAS.473.3615M} pointed out refined  criteria ($Q$ $<$ 1 for spiral arms, $Q$ $<$ 0.6 for fragmentation). 
They also found that a disk-to-star mass ratio is 0.5 or higher in the case of the unstable disk. 
These new criteria suggest that our observed $Q$ could be a little too high for the disks to be gravitationally unstable. 
The major uncertainty of $Q$ in this study originates from the uncertainty of the dust opacity index $\beta$, which is used for estimating the surface density $\Sigma$. 
The uncertainty from errors in the sound velocity $c_s$ and angular velocity $\Omega$ is limited, because these are proportional to the square root of a temperature or stellar mass. 
We note that $\beta$ could vary within 1-2, as discussed in Appendix \ref{app:dust}. 
Although we adopted $\beta$ of unity for conservative mass estimation; if $\beta$ = 2 is adopted, the dust mass opacity $\kappa_{\nu}$ becomes 40\% smaller. 
The total disk mass reaches 7 $M_{\sun}$ within 250 au (5 M$_{\sun}$ for the compact source within 100 au) in this case. 
We also obtain a 13\% lower dust temperature, because higher optical depth is expected at the center (see Appendix \ref{app:dust} again). 
This results in the averaged column density of 5.3 $\times$ 10$^{26}$ cm$^{-2}$ for the compact source and 2.4 $\times$ 10$^{25}$ cm$^{-2}$ for the resolved disk. 
Recalculated $Q$ values are 0.9 (average), 0.4 (minimum), and 2.0 (maximum) in the resolved disk. 
This is more consistent with the results of \citet{2018MNRAS.473.3615M}.

The spatial scale of the simulated disk in \citet{2018MNRAS.473.3615M} is quite similar to that of our observed disk, allowing us to directly compare disk structures.  
For example, we estimated the gas density of the disk, using the observed column density and disk scale height ($H$ $\sim$ $c_s/\Omega$). 
We obtained $H$ $\sim$13 au, at the radius of 100 au (see Appendix \ref{app:Q} for $c_s$  and $\Omega$ estimation). 
The gas density at 100 au is estimated as 1.7 $\times$ 10$^{-12}$ g cm$^{-3}$ for $\beta$ = 1 and 1.2  $\times$ 10$^{-11}$ g cm$^{-3}$ for $\beta$ = 2. 
The latter value \textcolor{red}{is comparable} to that reported in \citet{2018MNRAS.473.3615M} again. 
These comparisons could imply that \textcolor{red}{$\beta$ = 2} is a more adequate choice for G353; this will be examined by ongoing follow-up multi-band ALMA observations. 

Although $\beta$ uncertainty remains, the disk in G353 is the most self-gravitating disk currently known. 
We propose that this is the first detection of a gravitationally unstable accretion disk in high-mass star-formation. 
The obtained $Q$ values are significantly smaller than those estimated in previous studies toward high-mass star-forming regions \citep{2017A&A...603A..10B, 2018A&A...618A..46A}. 
This fact could be also explained by the young age of G353; i.e., the host object is still relatively less massive compared to other disk candidates, 
and also, the disk itself consists of the gas with a smaller angular momentum at the inner region of the natal core. 
It is evident that radiative heating from the 10 M$_\sun$ accreting star is insufficient to stabilize the young massive disk with the radius of 250 au. 
The highly time-dependent outflow activity \citep{2011MNRAS.417..238M, 2016PASJ...68...69M} may be related to the unstable nature of the accretion flow. 

The detected asymmetric structure can be also naturally explained by self-gravitating effects. 
The observed surface density contrast (up to 4-5; Fig. 3b) is easily produced by a spiral arm. 
If this is the case, the gravitational torque can redistribute angular momentum in the disk. 
Such an effect can produce further infalling streams within the inner region ($<$100 au) found in the previous study \citep{2017ApJ...849...23M}. 
The specific angular momentum of 2.4$\times$10$^{21}$ cm$^{2}$ s$^{-1}$ in the envelope is four times larger than that estimated for the inner streams.  
This indicates that 70\% of the angular momentum is removed in the disk, probably via the gravitational torque. 
A rapid mass accretion to the central star promoted by non-axisymmetric structures decreases the disk surface density; therefore, 
$Q$ should be adjusted to $Q$ $\sim$ 1-2 \citep[e.g.,][]{2017ApJ...835L..11T}, as seen in Fig. 1d.  

Another possible origin of such asymmetric structures is a tidal perturbation via a close-encounter event by a nearby cluster member \citep[e.g.,][]{2010ApJ...717..577T}. 
However, this is unlikely to be the case because no bright cluster member was detected in the ALMA FOV. 
Our continuum sensitivity corresponds to a 1$\sigma$ dust mass sensitivity of 8$\times$10$^{-3}$ M$_{\odot}$ beam$^{-1}$, adopting a dust temperature of 100 K. 
It is clear that G353 is the first high-mass cluster member in the natal cluster-forming clump, 
although there may be a very small dust condensation that could form a brown dwarf.

The clumpy local peaks seen in the western region could also trace the fragmentation process in the disk. 
Indeed, the outer disk radius of $\sim$250 au and the minimum $Q$ parameter of $\sim$ 1 are remarkably similar to that in the fragmented disk with the triplet low-mass protostar system in L1448 IRS3B \citep{2016Natur.538..483T}. 
These clumps may be the seed of a lower-mass binary companion, or accrete on the central object causing the accretion burst event \citep[e.g., ][]{2017MNRAS.464L..90M, 2018MNRAS.473.3615M, 2019MNRAS.482.5459M}. The latter event was recently detected in several objects \citep{2017NatPh..13..276C, 2018ApJ...854..170H}. 
The masses of two large clumps are 0.04 and 0.09 M$_\sun$ (Appendix D). 
These masses are for the $\beta$ = 1 case, and they are twice larger for the $\beta$ = 2 case. 

For comparison, we estimated the Jeans scale  ($\lambda_{\rm J}$) and Jeans mass ($M_{\rm J}$), in the disk.  
For the $\beta$ = 1 condition, the gas density is 1.6 $\times$ 10$^{-13}$ g cm$^{-3}$ and the temperature is $\sim$230 K at 250 au radius. 
The former was estimated from the averaged surface density of 1.0 $\times$ 10$^{25}$ cm$^{-2}$ in the eastern region, divided by the scale height of $\sim$43 au. 
We finally obtained $\lambda_{\rm J}$ of $\sim$104 au (0$\arcsec$.06) and $M_{\rm J}$ of $\sim$ 0.07 M$_{\odot}$.  
This is consistent with the observationally estimated masses of the clumps, and it is still comparable even for the $\beta$ = 2 condition, where $\lambda_{\rm J}$ and $M_{\rm J}$ decrease by 40\%. 
Although the estimated clump masses are 10-20 times larger than the accreting mass in the burst event in S255IR \citep{2017NatPh..13..276C}, it is consistent with the clump mass range formed by the disk fragmentation events in the numerical studies in \citet{2019MNRAS.482.5459M}.

Further high-resolution ($<$ 0\arcsec.05) ALMA observations are required, in order to directly investigate the innermost region. 
In this case, it is possible that the dust emission could be completely optically thick in a higher frequency than ALMA band 4. 
Lower-frequency interferometers such as the Square Kilometer Array or Next Generation Very Large Array may be able to resolve the innermost accretion process. 

\begin{figure*}[htb]
\epsscale{1.0}
\plotone{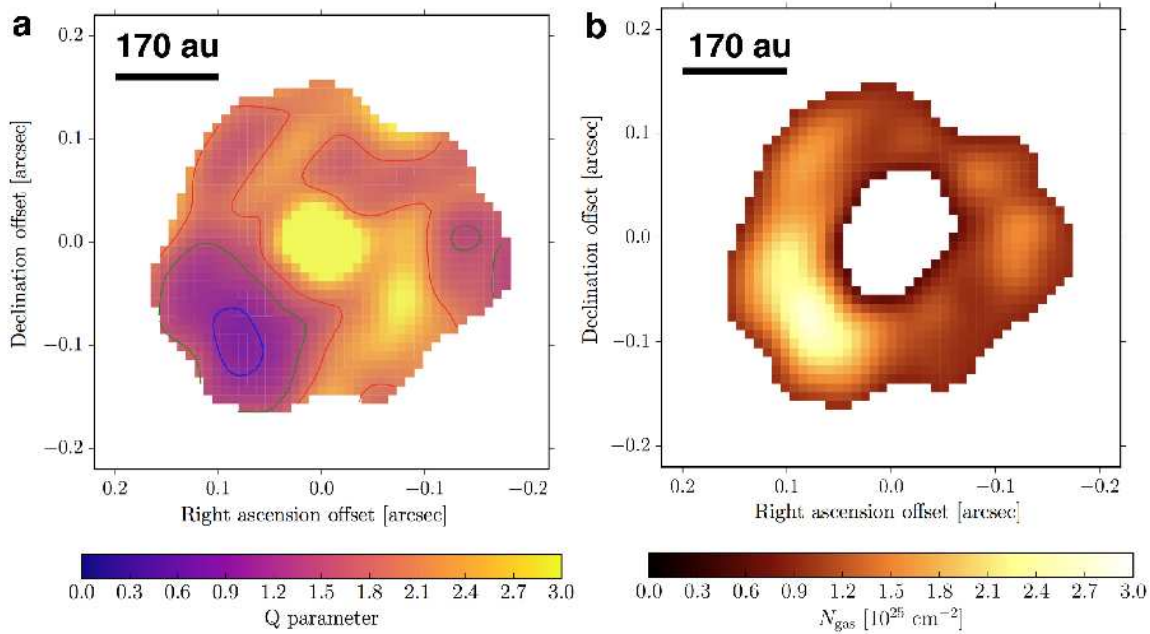}
\caption{(a):  Distribution of Toomre's $Q$ parameter. Blue, green, and red contours indicate $Q$ = 1.0, 1.5 and 2.0, respectively. 
The low brightness regions in Fig. 1b ($<$40 K) were masked and are shown in white. 
(b): Distribution of the surface density in the residual image. The minimum and maximum surface densities are 0.6 and 2.8, respectively. 
The low brightness regions ($<$40 K) are also shown in white. 
\label{fig:3}}
\end{figure*}

\section{Conclusions}
This study reports a new ALMA long baseline observation toward the nearly face-on accretion system around the high-mass protostellar object of 10 M$_{\sun}$. 
We have successfully resolved the infalling rotating envelope (5-7 M$_{\sun}$, 700 au in radius) and innermost centrifugal disk (2-7 M$_{\sun}$, 250 au in radius). 
The estimated envelope accretion rate suggests that the target object can be well younger  than 10$^4$ yr and still in the midpoint of its main accretion phase. 
The disk has the typical Toomre's $Q$ parameters of 1-2 at 100-250 au radius. 
The minimum  $Q$ reaches 0.4 under the largest dust opacity index of 2. 
This suggests that the disk in G353 is highly self-gravitating and gravitationally unstable. 
The observed non-axisymmetric disk structures are consistent with this picture. 
We found that 70\% of the initial angular momentum in the accretion flow could be removed via a gravitational effect such as a spiral arm. 
Our study suggests that the dynamical nature of a self-gravitating disk could dominate the early phase of high-mass star formation. 
This is remarkably consistent with the early evolutionary scenario of the low-mass protostar. 

This first bird's-eye view has also signaled the fact that the era of disk survey in high-mass star-formation has ended; its detailed structure and its diversities must be surveyed in the next decade. 
We emphasize that the viewing angle is critically important in order to study the innermost region of the massive disk. 
One must keep in mind the possibility that self-gravitating and unstable structure may exist, even in the cases of edge-on disks, apart from the stabilized and homogeneous Keplerian disk.

\acknowledgments
We thank A.J. Walsh and K. Tanaka for helpful comments and discussions in the observational strategy and analysis. 
We appreciate the anonymous referee's fruitful comments and suggestions. 
This work makes use of the ALMA dataset ADS$/$JAO.ALMA$\#$2016.1.01068.S. 
ALMA is a partnership of ESO (representing its member states), 
NSF (USA) and NINS (Japan), together with NRC (Canada), MOST and ASIAA (Taiwan), and KASI (Republic of Korea), in cooperation with the Republic of Chile. The Joint ALMA Observatory is operated by ESO, auI/NRAO and NAOJ. This work was financially supported by the MEXT/JSPS KAKENHI grant No. 15K17613 and 19H05082 (K.M.). 
K.M. was also supported by the ALMA Japan Research Grant of NAOJ ALMA Project, NAOJ-ALMA-219.
T.H. is financially supported by the MEXT/JSPS KAKENHI grant No. 16K05293, 17K05398, and 18H05222. 
S.T. acknowledges a grant from JSPS KAKENHI grant No. JP18K03703.
This work was supported by NAOJ ALMA Scientific Research Grant No. 2017-04A (S.T.).

\software{CASA \citep{2007ASPC..376..127M}, 
PyFITS \citep{1999ASPC..172..483B}, 
Matplotlib \citep{2007CSE.....9...90H}, 
} 

\facilities{ALMA}

\appendix
\section{Dust continuum emission}\label{app:dust}
In order to highlight the resolved circumstellar structure, 
we performed the elliptical Gaussian fitting for the compact dust continuum emission, and then, the best-fit Gaussian was subtracted from the original image (Table \ref{tab1}). 
The effect of this subtraction is less than 20 K at the midpoint of the residual ring in Fig. 1b. 
This does not change any conclusions in this paper. 
The spectral index of the compact component was determined using the total fluxes in Table \ref{tab1}. 

Most of the physical parameters of the dust continuum were estimated using the dust mass opacity at frequency $\nu$ ($\kappa_{\nu}$), dust opacity index $\beta$ and the gas-to-dust ratio ($R_{\rm gd}$). 
In this paper, we adopted $\kappa_{\nu}$ = 0.90$\times$($\nu$/230 GHz)$^{\beta}$ cm$^2$ g$^{-1}$, where we assumed the dust model coagulated in the dense gas (10$^6$ cm$^{-3}$) with the thin ice mantle \citep{1994A&A...291..943O}. 
Unless the typical size of the dust particle is very large as in the evolved protoplanetary disks \citep{2017ApJ...840...72L}, $\beta$ usually lies between 1 and 2 \citep[e.g.,][]{2016Natur.538..483T, 2017A&A...603A..10B}. 
We adopted $\beta$ = 1.0 in this work, and hence, $\kappa_{\rm 150 GHz}$ is to be 0.59  cm$^2$ g$^{-1}$. 
This is a conservative choice, i.e., higher $\beta$ results in a more massive disk with lower $Q$ value. 
It should be noted that $\beta$ $<$ 1.0 may be possible if the compact emission is optically thin, where the spectral index $\alpha$ is expressed as $\alpha$ $=$ 2$+\beta$. 
This is, however, clearly not the case, considering the absence of the CH$_{3}$OH lines toward the center. 

The averaged optical depth ($\tau_{\nu}$) and averaged dust temperature of the compact component ($T_{\rm ave}$) were determined by fitting the fluxes and source sizes at 45 and 150 GHz in Table \ref{tab1}. 
We used the Planck function and the relation $\tau_{\rm 150 GHz}$=$\tau_{\rm 45 GHz}$ (150 GHz$/$45 GHz)$^{\beta}$ in the fitting. 
The estimated $\tau_{\rm 150 GHz}$ and $T_{\rm ave}$ are 2.0 and 555 K, respectively. 
We note that any contribution from a free-free emission is negligible ($\sim$1 mJy or less) as suggested by \citet{2017ApJ...849...23M}. 
Next, obtained $\tau_{\rm 150 GHz}$ was converted to the surface density of the dusty gas as, 
\begin{eqnarray*}
\Sigma  =  \frac{\tau_\nu}{\kappa_{\nu}} R_{\rm gd}. 
\end{eqnarray*}
The total mass was, then, acquired by integrating $\Sigma$ over the source size. 

On the other hand, physical parameters in  the resolved structure were estimated by the following procedure. 
We first assumed the power-law profile of the dust temperature as, $T_{\rm dust}(r)$ = $T_{\rm 80 au}$($r$$/$80 au)$^{-0.4}$ K, where $r$ indicates a radius from the center. 
The referenced radius of 80 au is the averaged radius of the compact dust component. 
The assumed power-law index of -0.4 is a typical value for the embedded disk \citep[e.g.,][]{2015ApJ...813L..19J}. 
The referenced temperature $T_{\rm 80 au}$ was determined as 360 K, so as to match the averaged $T_{\rm dust}(r)$ inside 80 au and $T_{\rm ave}$. 
We, then, calculated $\tau_{\nu}$ from the ratio between $T_{\rm dust}$ and observed $T_{\rm B}$ in the resolved structure pixel-by-pixel as follows, 
\begin{eqnarray*}
\tau_{\nu} =  {\rm ln}\left(\frac{T_{\rm dust}}{T_{\rm dust}-T_{\rm B}}\right). 
\end{eqnarray*}
Finally, the obtained spatial profile of $\tau_{\nu}$ was converted to the surface density profile and total mass by using $\kappa_{\rm 150 GHz}$ and $R_{\rm gd}$ again. 
As mentioned in the main text, the dynamical heating effect by the non-axisymmetric disk structure can cause a deviation from the assumed $T_{\rm dust}(r)$ profile, 
i.e., spiral arms could have a higher temperature than surrounding region, because of gas compression. 
We clearly require further spatial resolution and/or multi-band dataset for discussing such a detailed condition.

\section{CH$_{3}$OH lines and infalling rotating envelope}\label{app:ch3oh}
Since all the detected CH$_3$OH lines showed similar spatial and kinematic profiles, we stacked multiple lines for better sensitivity. 
We divided detected lines into two categories based on the upper-state energy, i.e., "hot" transitions ($>$70 K) and "cold" transitions ($<$70 K). 
The stacking image and $PV$ diagram were made only for hot transitions (Fig. 2b), because the cold transitions showed a significant absorption feature in the blue-shifted outflow lobe. 
This outflow absorption will be reported in the forthcoming paper. 
The final image noise level (1 $\sigma$) was 1.1 mJy beam$^{-1}$ or 7.5 K. 
We adopted the cutoff signal-to-noise ratio of 10$\sigma$ ($\pm$ 75 K) for the integrated flux image. 
The regions outside this criterion were masked and shown by white color in Fig 2b. 
The $PV$ diagram (Fig. 2c) was made along the R.A. offset =0$\arcsec$, from north to south. 

We used the kinematic model of the infalling rotating envelope developed for the low-mass protostars \citep[e.g.,][]{2014Natur.507...78S, 2016ApJ...824...88O},which considers the conservation of energy and specific angular momentum. 
The rotational velocity ($\upsilon_{\rm rot}$)  and infall velocity ($\upsilon_{\rm inf}$) are expressed as, 
\begin{eqnarray*}
\upsilon_{\rm rot}  & = & \frac{l}{r_{\rm rot}} \\
\upsilon_{\rm inf}  & = & \sqrt{\frac{2GM_{\rm in}}{r} - \left(\frac{l}{r_{\rm rot}} \right)^{2}}, 
\end{eqnarray*}
where $G$ is the gravitational constant and $r_{\rm rot}$ is the projection of the radius $r$ to the envelope midplane. 
$M_{\rm in}$ is the enclosed mass that is fixed to 12 M$_{\sun}$ (10 M$_{\sun}$ for the star and 2 M$_{\sun}$ for the inner dusty disk). 
The specific angular momentum $l$  is also fixed to 2.4$\times$10$^{21}$ cm$^{2}$ s$^{-1}$, in order to have the centrifugal radius of 250 au that is the outer radius of the dusty disk. 
The rotational axis is set to be slightly inclined (8$^{\circ}$) to due east from the LOS as suggested by previous studies \citep{2016PASJ...68...69M, 2017ApJ...849...23M}. 
We assume that the envelope is rotating counterclockwise and purely axisymmetric. 
The latter is adequate along the north--south direction, although the east--west asymmetry is clear in Fig. 2b.  
The envelope becomes geometrically thicker outwards and the flaring angle is set to be constant for simplicity. 
We did not consider any outflow contribution in the north--south direction, since the compact water maser jet in similar spatial scale was clearly along the east--west direction \citep[e.g.,][]{2016PASJ...68...69M, 2017ApJ...849...23M}.

\section{Toomre's $Q$ parameter}\label{app:Q}
Toomre's $Q$ parameter is expressed as, 
\begin{eqnarray*}
Q = \frac{c_{\rm s}\kappa}{\pi G\Sigma}, 
\end{eqnarray*}
where $c_{\rm s}$ is the sound velocity, $\kappa$ is the epicyclic frequency, $\Sigma$ is the surface density. 
We calculated $c_{\rm s}$ from the $T_{\rm dust}(r)$. 
The profile of $\Sigma$ is already given by the analysis above. 
$k$ is given by the angular velocity assuming that the dusty disk is in the Keplerian rotation around 10 M$_{\sun}$ star. 

\section{Clump mass}\label{app:clump}
We defined two large clumps in the west by the surface density above 1.0$\times$10$^{25}$ cm$^{-2}$ that is the average  surface density in the western region. 
The clump masses were estimated to be 0.04 and 0.09 M$_{\sun}$, integrating the surface densities.

\end{document}